\documentstyle[aps,prl,twocolumn,epsfig]{revtex}
\begin{document}
\bibliographystyle{prsty}
 \wideabs{
\title{Parametric Resonance in the Drift Motion
of an Ionic Bubble in near critical Ar Gas}
\author{A.F.Borghesani\cite{byline}
 and F.Tamburini\cite{byline2}}
\address{Istituto Nazionale per la Fisica della Materia\\
Department of Physics ``G.Galilei'', via F. Marzolo, 8, 
I--35131 Padua, Italy}
 \maketitle
\begin{abstract}
The drift mobility of the O$_{2}^{-}$ ion in dense Argon gas near the 
 liquid--vapor 
critical point has been measured as a function of the 
density. Near $T_{c}$
the zero--field density--normalized
ion mobility $\mu_{0}N$ shows a deep minimum at a density 
smaller than $N_{c}.$
This phenomenon was previously 
intepreted as the result of the formation of a correlated cluster 
of Argon atoms around the ion because of the strong electrostriction 
exerted by the ion on the highly polarizable and compressible gas. 
We now suggest that a possible alternative explanation is related 
to the onset of a parametric resonance of a bubble surrounding the 
ion. At resonance a large amount of energy is dissipated by sound 
waves in addition to viscous dissipation processes, 
resulting in the large mobility drop observed. 
\end {abstract}}
\noindent The transport properties of negative ions in dense rare gases and
liquids have been recently subject of renewed interest because they can give
information on how the microscopic structure of the fluid around the ion is
modified by the ion--atom interaction \cite{kh1,borg93}. Moreover, 
the possibility to greatly change with relative
ease the gas density $N$ close to 
the critical point of the liquid--vapor transition gives the experimenters the
unique opportunity to investigate the transition from the kinetic to the
hydrodynamic transport regime \cite{borg95}.

The drift mobility of O$_{2}^{-}$ in dense Ne gas \cite{borg93} or in liquid
Xe \cite{hsk94} can be satisfactorily explained in terms of hydrodynamics if
ions are assumed to be surrounded by a complex structure. The competition
between short--range repulsive exchange forces between the weakly bound
outer electron of the ion and the electrons of the atoms and the long--range
polarization attraction of the ion with the atoms 
leads to the formation of a microcavity around the ion, which is in turn
surrounded by a strong density enhancement \cite{kh1,kh2}.

A self--consistent field model has been developed to compute the structure
shape \cite{kh2}. The size of the complex is taken as effective hydrodynamic
radius $R$ of the ion and its drift mobility is calculated by means of the
Stokes formula $\mu_{0}=e/6\pi\eta R,$ where $\eta$ is the viscosity. The
agreement with O$_{2}^{-}$ mobility data in liquid Xe \cite{hsk94} is quite
good, but the data in Ne gas near 
$T_{c}$\cite{borg93} are almost quantitatively reproduced only for $N> 
N_{c}.$ 
In particular, the model does not 
reproduce even qualitatively the little dimple in the density--normalized
mobility $\mu_{0}N$ for $N\approx 0.7 N_{c} .$ In this Letter we show that
this general feature of the mobility \cite{cantelli} can be explained by
assuming that under certain termodynamic conditions a further dissipation
mechanism becomes active in addition to viscous processes, namely sound wave
emission by oscillations of the structure surrounding the ion.

Since the influence of electrostriction on the structure formation 
depends on the gas polarizability, we have carried out O$_{2}^{-}$ mobility
measurements in Ar gas \cite{borg97b} because its polarizability is much
larger than that of Ne. We used the pulsed photoinjection technique as for
Ne \cite{borg93} and He \cite{borg95}. Details of the experiment are
reported in literature \cite{borg88,borg90}. A small bunch of electrons is
injected into the gas by irradiating a photocathode by means of a short UV
light pulse. Electrons are captured by O$_{2}$ impurities (in a
concentration of $10-100$ {\sl p.p.m.}) forming stable O$_{2}^{-}$ ions \cite
{borg97}. They drift under the action of an external electric field $E$
towards the anode inducing a detectable current. The analysis of its time
dependence allows the determination of the drift time and, hence, of the
drift velocity $v_{D}$ \cite{borg90b}. Finally, the mobility is calculated
as $\mu = v_{D}/E.$ 
\begin{figure}[htbp]
\epsfig{file=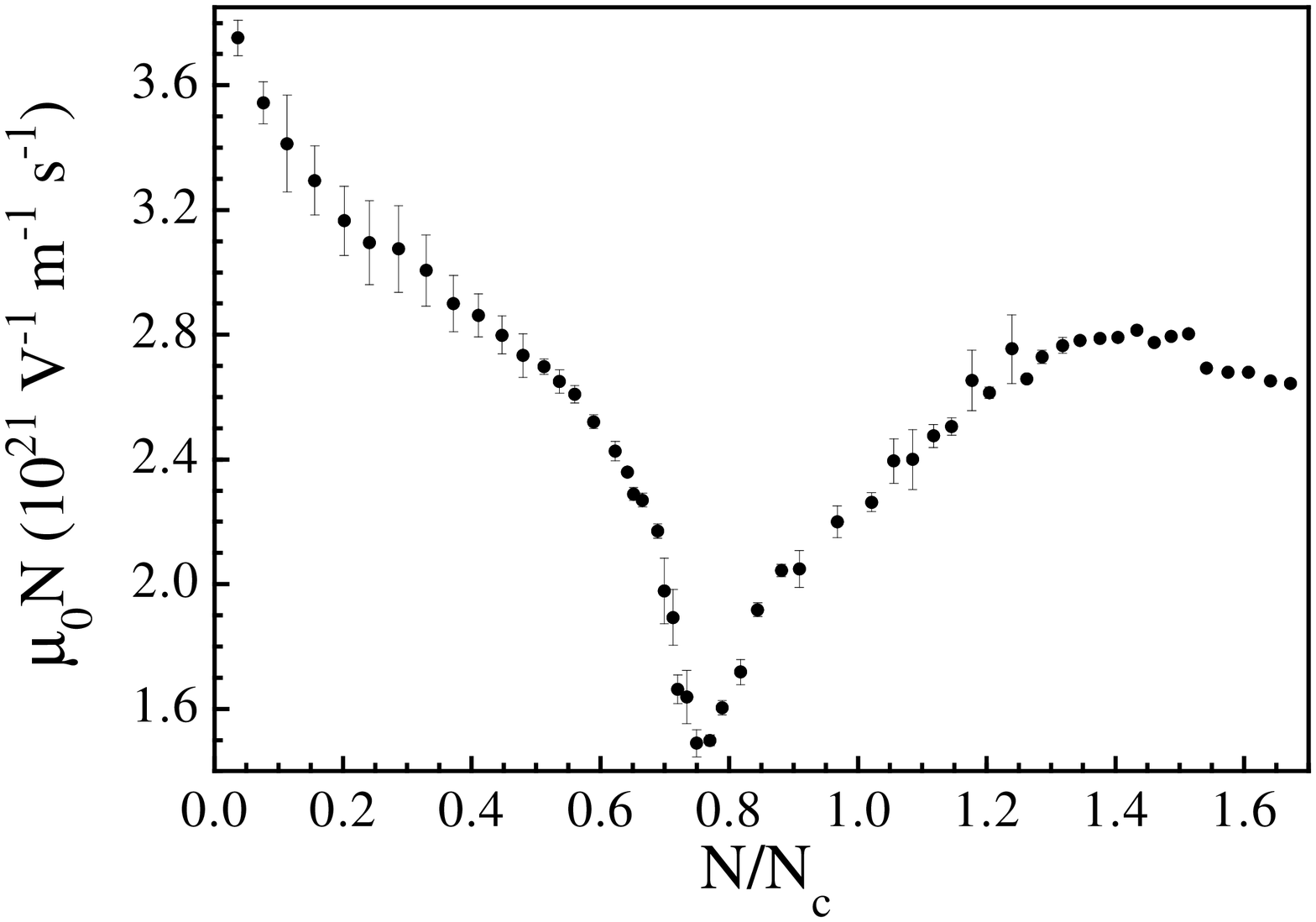,width=\columnwidth}
\caption{\small Experimental zero--field density--normalized 
mobility             
$\mu_{0}N$ vs. $N/N_{c}$ for $T=151.5\,\mathrm{K}\, (T/T_{c}\approx 1.005).$ }
\label{fig1}
\end{figure}
\noindent In Fig.\ref{fig1} we show the measured zero--field
density--normalized mobility $\mu_{0}N$ vs. the reduced density $N/N_{c}$
for $T=151.5 \, \mathrm{K }\, (T/T_{c}\approx 1.005)$ ($N_{c}=8.08\, 
\mathrm{atoms}\cdot\mathrm{nm}^{-3}$ $T_{c} = 151.7\, \mathrm{K}$). $
\mu_{0}N $ shows a very deep minimum for $N/N_{c}\approx 0.76.$ The 
Stokes formula with constant radius reproduces the data, at most, for $
N/N_{c}\ge 1.3$ \cite{borg97b}. The data are heuristically explained by
assuming that a cluster of Ar atoms forms around the ion as a consequence of
electrostriction \cite{borg97b}. This large structure interacts
hydrodynamically with the gas. 
The effective hydrodynamic radius 
depends on the local properties of the fluid. The 
good agreement with the data for $N/N_{c}\ge 0.5$ is obtained by introducing
into the effective radius an adjustable contribution proportional to the
size of the strongly correlated cluster. Thus, the 
hydrodynamic radius shows a complicated {\it ad hoc} density dependence,
whose meaning is not easy to grasp. The reason might be that this model
focuses only on the structure determined by the long--range part of the
ion--atom interaction potential and does not take into account the presence
of the microcavity closely surrounding the ion. Therefore, it neglects the
possibility that the cavity may oscillate. In this case the energy of the moving complex,  ion plus
structure or, briefly, {\it ionic bubble}, can be dissipated by sound wave
radiation in addition to viscous processes.

The present problem bears many affinities with that of the dynamics of
electron bubble formation in liquid He \cite{rosenblit95}, the adiabatic
process where the repulsive electron--atom interaction displaces a large
number of atoms far away from the electron which is localized within the
fluid dilation, or the electron solvation dynamics in water, where cavity
contraction is induced by long--range attractive polarization interactions 
\cite{rips97}. These problems have been treated in terms of a hydrodynamic
model of cavity expansion \cite{rosenblit95} or collapse \cite{rips97}. The
dynamics of the cavity boundary is determined by the flow of the surrounding
solvent. This is assumed to be spherically symmetric with velocity in the
radial direction. In absence of energy dissipation and by assuming the
incompressibility of the solvent, the cavity boundary velocity $U$ is
described by the Rayleigh--Plesset (RP) equation 
\begin{equation}
\label{eq:rp1}{\frac{{\mathrm{d}}U}{{\mathrm{d}}t}}=-{{{3U^2}\over 2R}}+{
{1\over \rho R}}\left[ p(R)-p_e\right] 
\end{equation}
where $R$ is the time dependent cavity radius, $\rho $ is the gas mass
density, $p_e$ is the external pressure, and $p(R)$ is the pressure on the
cavity boundary, expressed in terms of the free energy change $F$ in the
process: $p(R)=-\left( \partial F/\partial R\right) /4\pi R^2.$ The
equilibrium radius is obtained by minimizing $F,$ but in absence of
dissipation the system will not equilibrate. The cavity oscillates around
the equilibrium value of the radius. Energy dissipation causes the damping
of the oscillations. Within the given approximations, the first passage
time, i.e., the time required for the cavity radius to expand (contract)
from its initial value to its maximum (minimum) is taken as the lower bound
for the relaxation timescale of the process. Viscosity and emission of sound
waves drive the system towards the equilibrium \cite{rosenblit95,rips97}.

It is obvious that also the drift motion of the ion is affected if energy is
dissipated also by emission of sound waves. We therefore address the issue
of the motion of the ionic bubble in dense Ar gas, considering the
oscillations of the bubble boundary. We assume a spherical cavity with
well--defined boundary of radius $R(t)$ with equilibrium value $R_0.$ The
main contribution to the restoring force is due to the surface tension.
Although a gas has no surface tension, 
the enhanced density region around the cavity acts 
as a well--defined interface.  Oscillations are initiated by 
collisions with energetic Ar atoms. Owing to the bubble smallness and 
to the asimmetry induced by the average motion driven by the external electric 
field, 
these collisions are not very frequent and occur in a  spatially non 
homogeneous way.
The equation of motion of the bubble boundary
is then given by the RP equation, and, beside the impulsive nature of the
driving force, is the same equation used for the bubble dynamics in
the sonoluminescence experiments \cite{brenner95}. 

So, for small distortion of the spherical shape a solution for the bubble
radius is sought in the form $R_0+a_nY_n,$ where $Y_n$ is a spherical
harmonic of degree $n,$ and $a_n$ are the distortion amplitude coefficients.
As bubble oscillations are initiated by collisions with energetic Ar atoms,
this phenomenon can be basically described by a  kicked rotator model, with
the following equation of motion for the simple normalized periodic case: $ 
a_n^{\prime \prime }+\Gamma a_n^{\prime }-Kf(a_n)\sum_{m=0}^\infty \delta
(t-m\tau )=0$, where $m$ is an integer, $\Gamma $ the damping constant and $
\tau $ the period between two kicks. Imposing a stochastic behaviour to the
potential function, we can model collision events occurring with a gaussian
distribution, e.g. in the quantum kicked rotator and electron localization
problem \cite{Fishman82}. All properties of this equation are well known
results of nonlinear dynamical system theory, which assure, under certain
assumptions, the existence of a universal route to chaos. % as pitchfork
%bifurcations. % (and the correspondence of this equation, in the case of strong
%damping, to the logistic map).
In
the limit of small forcing the dynamics of the distortion amplitude 
can be cast in the form of the Mathieu--Hill (MH) like
equation \cite{brenner95} 
\begin{equation}
\label{eq:mathil}b_n^{\prime \prime }+2\xi _mb_n^{\prime }+\omega _m^2\left(
1+\epsilon _m\cos {2\tilde t}\right) b_n=0
\end{equation}
with $b_n\propto R_0^{3/2}a_n(t).$ $\omega _m^2=(\omega _0/\omega )^2$ is
the square of the ratio between the natural frequency of the bubble, $\omega
_0,$ and the excitation frequency $\omega .$ The natural frequency is given
by $\omega _0^2=\beta _n\sigma /\rho R_0^3,$ where $\beta _n=(n-1)(n+1)(n+2),
$ $\sigma $ is the surface tension, and $\rho $ is the mass density of the
gas \cite{brenner95}. Primes denote differentiation with respect to the
dimensionless time $\tilde t=\omega t.$ Assuming that the bubble is empty,
the surface tension can be calculated by using the parachoric formula \cite
{hirsch} $\sigma =(PN)^4,$ where $N$ is the gas number density and $P$ is a
constant. For Argon $P\approx 1.39\times 10^{-29}\,\mathrm{J}^{1/4}\, 
\mathrm{m}^{5/2}.$ The term $\xi _m=2n(n+2)\eta /\rho \omega R_0^2$ is a
damping coefficient related to the viscosity $\eta $ whose values are found
in literature\cite{vangulik}. Of the forcing term only the first Fourier
component of amplitude $\epsilon _m$ has been retained.

For a dissipation--free system the Mathieu--Hill equation is
known to give origin to parametric instability, when deviations from the
spherical shape accumulate over many oscillation cycles. Floquet's theorem
states that solutions of Eq.\ref{eq:mathil} take the form \cite{ince} 
\begin{equation}
\label{eq:floq}b_n\left( \tilde t\right) =e^{\mu _f\tilde t}P_n\left( \tilde 
t\right) 
\end{equation}
where $\mu _f$ is the Floquet's index and $P_n$ is periodic. 
$\omega _m$ and $\epsilon _m$ span a plane geometrically divided into
stability and instability regions\cite{berge}. We 
focus on the $n=2$ mode.
In the instability regions $\mu _f^2>0$ and the envelope of the solutions to
the MH equation grows exponentially. In the physical systems at hand this
amplitude cannot grow indefinitely because dissipation stabilizes the
surface dynamics. In the stability regions where ${\tt Re}\,\mu _f=0$ and $ 
{\tt Im}\,\mu _f=2\omega _m$ the solutions are periodic with angular
frequency $\omega _0$ with a small modulation at angular frequency $2\omega
_0.$ In this case the system experiences parametric resonance.
In the limit of small forcing $ 
(\epsilon _m\rightarrow 0),$ in absence of dissipation $(\xi _m\rightarrow 0)
$ and close to the resonance $(\omega _m\approx 1),$ the non--zero solution
for the Floquet's index is $\mu _f\approx 2i\omega _m=2i{\omega 
_0/\omega } $ \cite{berge}.
We now show that ${\tt Im}\,\mu _f$ is related to the measured mobility $\mu
_{0.}$ From the expression of the bubble eigenfrequency, using the
parachoric formula, we can write 
\begin{equation}
\label{eq:omemme}\omega _m={\frac g\omega }N^{3/2}
\end{equation}
where $g=(N_A\beta _nP^4/MR_0^{3/2})^{1/2}$ is a constant. $N_A$ is the
Avogadro's number and $M$ is the atomic weight of Argon. $\omega $ is the
excitation frequency. Therefore, under the hypotesis of collisional induced
oscillations, $\omega =\omega _{coll}/l=2\pi \nu _{coll}/l,$ where $\nu
_{coll}$ is the collision frequency and $l$ defines the order of a suitable
subharmonic. In the Knudsen regime the mobility of a heavy ion of radius $R_0
$ scattered off light particles of mass $m$ is 
\begin{equation}
\label{eq:mobiknudsen}\mu _0N={\frac{3e}{8R_0^2\sqrt{2\pi mk_{\mathrm{B}}T}}}
\end{equation}
and is related to the collision frequency by \cite{kh1} 
\begin{equation}
\label{eq:nucollmobilita}\nu _{coll}={\frac{3eN}{m\left( \mu _0N\right) }}
\end{equation}
By inserting this result into Eq.\ref{eq:omemme} we get 
\begin{equation}
\label{eq:mufmu0n}{\tt Im}\,\mu _f\propto \sqrt{N}\left( \mu _0N\right) 
\end{equation}
where $\mu _0N$ is the experimentally measured zero--field
density--normalized mobility. Therefore, $\sqrt{N}\mu _0N$ can be used to
make a stability analysis of the bubble boundary motion.

In the real system, the ideal conditions of small forcing and absence of
dissipation are not met and a parametric resonance is characterized by a
minimum of the effective Floquet index, approximated by ${\tt Re}\, 
\mu_{f},$ which is shown to be related to $\sqrt{N}\mu_{0}N.$ The
minimum observed for $N/N_{c}\approx 0.76 \> (N\approx 6.2\>\mathrm{atoms}
\cdot\mathrm{nm}^{3})$ is due to parametric resonance of the ionic bubble.
We follow a heuristic line of reasoning. Once excited, the bubble starts
oscillating with characteristic frequency ${\tt Im}\{ \mu_{f}\}$ and growing
amplitude given by $A(t)=\exp{({\tt Re}\{ \mu_{f}\}t)}.$ The growth of the
amplitude is limited because collision processes can absorb the oscillation
energy. Therefore, if ${\mathcal{T}} $ is the time interval during which the
bubble oscillates, ${\tt Re}\{ \mu_{f}\} {\mathcal{T}}\simeq {\mathcal{O}}
(1).$ On the other hand, according to calculations in the cavitation model 
\cite{rips97}, the oscillations last for a few cycles, hence ${\tt Im}\{
\mu_{f}\} {\mathcal{T}} \simeq {\mathcal{O}}(1).$ Then, we have ${\tt Re}\{
\mu_{f}\}\simeq {\tt Im}\{ \mu_{f}\} $ and, owing to Eq.\ref{eq:mufmu0n}, $
\sqrt{N}(\mu_{0}N)$ is also related to the real part of the Floquet's index.
The {\sl Ansatz} must be then verified by a direct numerical stability analysis
of Eq.\ref{eq:mathil}.

\begin{figure}[htbp]
 \epsfig{file=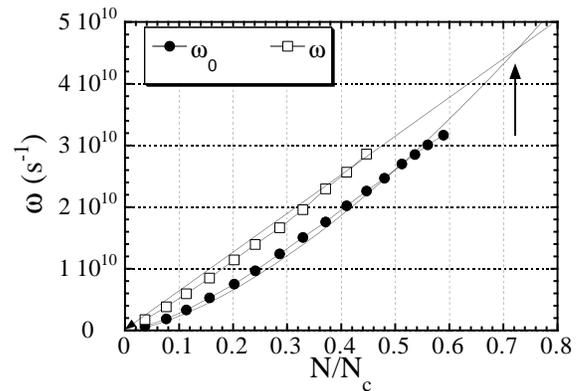,width=\columnwidth}
 \protect\caption{\small Excitation frequency $\omega$ and 
 eigenfrequency $\omega_{0}$ of the oscillating ionic bubble.}
 \label{fig2}
 \end{figure}

The minimum of $\mu_{0}N$ occurs at a density $N/N_{c}\approx 0.76$ where
the resonance condition $\omega=\omega_{0}$ is met. In Fig.\ref{fig2} we
plot vs. $N$ both $\omega_{0}\propto N^{3/2}$ and $\omega=\omega_{coll}/l
\propto N, $ approximately. 
The two curves intersect for the right density value if $l\approx 2^{11}.$
So, the excitation frequency must be $\approx 2^{11}$ times lower than the
average collision frequency given by Eq.\ref{eq:nucollmobilita}. The factor $
2^{11}$ can be explained assuming that only collisions with Ar atoms in the
tail of the Maxwell--Boltzmann energy distribution function, namely those
with average kinetic energy in excess of $3k_{\mathrm{B}}T,$ are energetic
enough to initiate bubble oscillations. These energetic collisions are less
frequent than average by the right factor.

\begin{figure}[htbp]
\epsfig{file=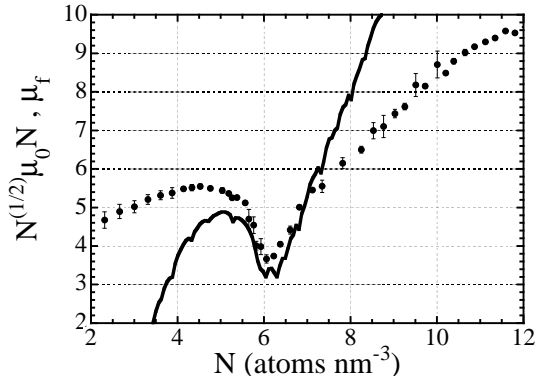,width=\columnwidth}
\caption{\small Floquet's index $\mu_{f}$ obtained by numerical 
integrating the MH equation with $R_{0}=10\,{\mathrm{\AA}}$ and 
$\epsilon_{m}=1100$ compared to the experimental 
data $\mu_{0}N\sqrt{N}$ vs. $N/N_{c}.$}
\label{fig3}
\end{figure}

We have numerically integrated the MH equation (Eq.\ref{eq:mathil}) by
assuming an equilibrium radius $R_{0}\approx 10\, \mathrm{\AA},$ and have
carried out the usual analysis of the stability bands. Usually, the
Floquet's index is plotted vs. $\omega_{m}^{2}.$ However, to compare the
experimental data with the results of numerical analysis we have converted $
\omega_{m}$ to $N$ by means of Eqns. \ref{eq:omemme} and \ref{eq:nucollmobilita}. 
In Fig. \ref{fig3} we plot $(\mu_{0}N)\sqrt{N}$ and 
${\tt Re}\, \mu_{f}$ as a function of the gas density $N.$ The Floquet's
index shows a number of small bands, but also a very deep minimum at the
same $N$ of the experimental $\mu_{0}N.$ This minimum is therefore
associated with a strong parametric resonance of the ionic bubble. The
larger width of the experimental data can be explained by considering the
fact that the experimental result is an average over a distribution of ionic
bubble radii. At the condition of parametric resonance the amplitude of
oscillation can reach substantial values and the kinetic energy gained by
the external electric field can be efficiently dissipated by emission of
sound waves. 
\noindent Close to the critical point the sound velocity $c$ is minimum ($
\approx 190 \, m/s$\cite{tegeler}) and sound dissipation is favoured. The
sound 
intensity irradiated by an oscillating bubble 
in the long wavelength limit is \cite{landau} $I = 2\pi \rho
c\omega^{2}R_{0}^{4}.$ At the density of the minimum $\mu_{0}N$ the 
intensity irradiated per cycle is in the tens of meV range and a relevant
part of the energy gained by the ion from the electric field 
is dissipated by means of this process in addition to the usual viscous
processes. At higher $N$ the stiffness of the bubble surface becomes too
large and the oscillations cannot be initiated so easily as in the resonance
region. Hence, at high $N$ the contribution of the sound to energy
dissipation is limited and $\mu_{0}$ is determined by the viscosity. This is
why the high$-N$ region $\mu_{0}N$ can be well described by the Stokes
formula. On the contrary, at much lower $N$ bubbles cannot form and $\mu_{0}N
$ is determined by dissipation processes other than sound  
emission.

In the region where stable bubbles 
exist a further 
dissipation mechanism, though probably very small, should be considered.
Within the bubble the ion undergoes a chaotic motion bouncing back and forth
from the inner bubble wall. Molecular Dynamics studies \cite{poll} have shown a vibration of the ion 
within the cavity. 
The ion should therefore behave as an emitting antenna of 
characteristic frequency 
$\omega_{e} \approx 2\pi v_{th}/2R_{0},$ where $v_{th}=(3k_{\mathrm{B}
}T/m_{i})^{1/2}$ is the ion thermal velocity and $m_{i} $ its mass. At 
the experiment
temperature $\omega_{e}\approx 10^{12}\, rad\, s^{-1}.$
Moreover, this radiation should be modulated by the slower oscillation of
the bubble boundary.
We finally note that in this problem there is a moving interface between
media of different polarizability, crossed by the strong electrical field of
the ion. Therefore, an experiment could be designed to detect the expected
quantum radiation emitted as a dynamic Casimir effect, that some authors
consider the physical cause of sonoluminescence\cite{eber}.
\vspace*{-2ex}
% \begin{thebibliography}{99}

% \end{thebibliography}


\begin{references}
\vspace*{-8ex}

\bibitem[*]{byline}e-mail address: borghesani@padova.infm.it

\bibitem[+]{byline2}e-mail address: tamburini@iol.it

\bibitem{kh1}  K.F.Volykhin, A.G.Khrapak, and W.F.Schmidt, J.E.T.P. {\bf 81}, 
901 (1995)

\bibitem{borg93}  A.F.Borghesani, D.Neri, and M.Santini, Phys.Rev. {\bf E 48}, 
1379 (1993)

\bibitem{borg95}  A.F.Borghesani, F.Chiminello, D.Neri, and M.Santini, Int.
J. Thermophys. {\bf 16}, 1235 (1995)

\bibitem{hsk94}  O.Hilt, W.F.Schmidt, and A.G.Khrapak, IEEE Trans. Dielect.
Electr. Ins. {\bf 1}, 648 (1994)

\bibitem{kh2}  A.G.Khrapak and K.F.Volykhin, J.E.T.P., {\bf 88}, 320 (1999)

\bibitem{cantelli}  R.Cantelli, I.Modena, and F.P.Ricci, Phys. Rev. {\bf 171}
, 236 (1968)

\bibitem{borg97b}  A.F.Borghesani, D.Neri, and A.Barbarotto, Chem. Phys.
Lett. {\bf 267}, 11 (1997)

\bibitem{borg88}  A.F.Borghesani, L.Bruschi, M.Santini, and G. Torzo, Phys.
Rev. {\bf A 37}, 4828 (1988)

\bibitem{borg90}  A.F.Borghesani and M.Santini, Phys. Rev. {\bf A 42}, 7377
(1990)

\bibitem{borg97}  Dino Neri, A.F.Borghesani, and M.Santini, Phys. Rev. {\bf 
E 56}, 2137 (1997)

\bibitem{borg90b}  A.F.Borghesani and M.Santini, Meas. Sci. Technol. {\bf 1}
, 939 (1990)

\bibitem{rosenblit95}  M.Rosenblit and J.Jortner, Phys. Rev. Lett. {\bf 75},
4079 (1995)

\bibitem{rips97}  I.Rips, J. Chem. Phys. {\bf 106}, 2702 (1997)

\bibitem{Fishman82}  S. Fishman, D.R. Grempel, and R.E. Prange, Phys. Rev.
Lett. {\bf 49}, 509, (1982);F.M. Izraelev and D.L. Shepelanskii, Theor.
Math. Phys. {\bf 43}, 417 (1980); H. G. Shuster, {\sl Deterministic Chaos},
(VCH, Weinheim, 1988)

\bibitem{brenner95}  M.P.Brenner, D.Lohse, and T.F.Dupont, Phys. Rev. Lett., 
{\bf 75}, 954 (1995)

\bibitem{hirsch}  J.O.Hirschfelder, C.F.Curtiss, and R.B.Bird, {\sl 
Molecular Theory of Gases and Liquids} (Wiley, New York, 1964)

\bibitem{vangulik}  N.J.Trappeniers, P.S. van der Gulik, and H. van den
Hooff, Chem. Phys. Lett., {\bf 70}, 438 (1980)

\bibitem{ince}  E.L.Ince, {\sl Ordinary Differential Equations} (Dover, New
York, 1956)

\bibitem{berge}  P.Berg\`e, Y.Pomeau, and C.Vidal, {\sl Order and Chaos}
(Wiley, New York, 1984)

\bibitem{tegeler}  C. Tegeler, R. Span, and W. Wagner, VDI
Fortschritt-Berichte, Reihe 3, Nr. 480, VDI Verlag, D\H usseldorf (1997).

\bibitem{landau}  L.D.Landau and E.M.Lifshitz, {\sl Fluid Mechanics}, 74
(Pergamon, Oxford, 1987)

\bibitem{poll}  E.L.Pollock and B.J.Alder, Phys. Rev. Lett. {\bf 41}, 903
(1978)

\bibitem{eber}  C.Eberlein, Phys. Rev. Lett. {\bf 76}, 3842 (1996) 

\end{references}
\end{document}